\documentclass[aps,prl,twocolumn,superscriptaddress,preprintnumbers,showpacs]{revtex4}
\usepackage{graphics}
\begin{document}


\title{The $Z\,\rightarrow \,\gamma  \gamma,\;gg$ 
Decays in the Noncommutative Standard Model}

\author{W.~Behr}
\affiliation{Theoretische Physik, Universit\"at M\"unchen, 
Theresienstr. 37, 80333 M\"unchen, Germany}
\email[]{wess@theorie.physik.uni-muenchen.de}
\author{N.G.~Deshpande}
\affiliation{Institute of Theoretical Science, University of Oregon, 
Eugene, OR94703, USA}
\email[]{desh@oregon.uoregon.edu}
\author{G.~Duplan\v{c}i\'{c}}
\affiliation{Theoretical Physics Division, 
Rudjer Bo\v{s}kovi\'{c} Institute, P.O.Box 180,
10002 Zagreb, Croatia}
\email[]{gorand@thphys.irb.hr}
\author{P.~Schupp}
\affiliation{International University Bremen, Campus Ring 1, 28759 Bremen, Germany}
\email[]{p.schupp@iu-bremen.de}
\author{J.~Trampeti\' c}
\affiliation{Theoretical Physics Division, 
Rudjer Bo\v{s}kovi\'{c} Institute, P.O.Box 180,
10002 Zagreb, Croatia}
\affiliation{Theory Division, CERN, CH-1211, Geneva 23, Switzerland}
\email[]{josip.trampetic@cern.ch}
\author{J.~Wess}
\affiliation{Theoretische Physik, Universit\"at M\"unchen, 
Theresienstr. 37, 80333 M\"unchen, Germany}
\affiliation{Max-Planck-Institut f\"ur Physik, F\"ohringer Ring 6, 
80805 M\"unchen, Germany}

\date{\today}

\begin{abstract}
On noncommutative spacetime, the Standard Model (SM) allows
new, usually SM forbidden, triple gauge boson interactions.
In this letter we propose the SM strictly forbidden $Z\rightarrow \gamma\gamma$ and $Z\rightarrow gg$
decay modes coming from the gauge sector of the Noncommutative Standard Model (NCSM) as a 
place where noncommutativity could be experimentally discovered.
\end{abstract}

\pacs{12.60.Cn, 13.38.Dg, 02.40.Gh}

\maketitle


In this article we consider strictly SM forbidden decays
coming from the gauge sector of the NCSM
which could be probed in high energy collider experiments.
This sector is particularly 
interesting from the theoretical point of view. 
It is the place where different models show the greatest differences.
In particular there are models that do not require any new triple 
gauge boson interactions. 
This depends
on a choice of representation.
It is, however, important to emphasize that generically one should expect
triple boson interactions. 
We will in particular argue that a model that does have 
new triple gauge boson interactions is natural as an effective theory
of noncommutativity.
Our main results are summarized in equations (\ref{eqn1}) to (\ref{eqn3}).

The idea that coordinates may not commute 
can be traced back to Heisenberg.
A simple way to introduce a noncommutative structure into
spacetime is to promote the usual spacetime 
coordinates $x$ to noncommutative (NC) coordinates 
$\hat x$ with~\cite{sny}
\begin{equation}
\left[{\hat x}^{\mu},{\hat x}^{\nu} \right]=i\theta^{\mu\nu}, \quad
\left[\theta^{\mu\nu},{\hat x}^{\rho} \right]=0,                \label{CR}
\end{equation}
were $\theta^{\mu\nu}$ is a constant, real, antisymmetric matrix.
The noncommutativity scale $\Lambda_{NC}$ is fixed
by choosing dimensionlfess matrix elements 
$c^{\mu\nu}=\Lambda_{NC}^2\,
\theta^{\mu\nu}$ of order one.
The original motivation to study such a scenario
was the hope that the introduction
of a fundamental scale
could deal with the infinities of
quantum field theory in a natural way.
The simple commutation relation (\ref{CR})
with constant $\theta^{\mu\nu}$ 
fails to provide a complete regularization~\cite{Filk},
but more complicated noncommutative structures
can indeed introduce spacetime lattice structures into the
theory that are compatible with a deformation of \emph{continuous} 
spacetime symmetries (see, e.g., \cite{Cerchiai}). This is in contrast
to the situation in ordinary lattice
field theory, where only discrete translation symmetries survive.
Aside from these technical merits, the possibility of 
a noncommutative structure 
of spacetime is of interest in its own right
and its experimental discovery would be a result of fundamental importance. 

Noncommutative gauge theory has become a focus of interest
in string theory and M-theory with the work given in Ref.\cite{CDS}. 
Noncommutativity of spacetime
is very natural in string theory and can be
understood as an effect of the interplay of closed and open strings.
The commutation relation (\ref{CR}) 
enters in string theory through
the Moyal-Weyl star product
\begin{equation}
f \star g = \sum_{n=0}^\infty \frac{\theta^{\mu_1 \nu_1} 
\cdots \theta^{\mu_n \nu_n}}{(-2i)^nn!}  
\partial_{\mu_1}\ldots\partial_{\mu_n} f
\cdot\partial_{\nu_1}\ldots\partial_{\nu_n} g.
\end{equation} 
For coordinates: $x^\mu \star x^\nu - x^\nu \star x^\mu = i \theta^{\mu\nu}$.
The tensor $\theta^{\mu\nu}$ is determined by a NS  $B^{\mu\nu}$-field and
the open string metric $G^{\mu\nu}$~\cite{DH}, which both
depend on a given closed string background. 
The effective physics on D-branes is most naturally captured by
noncommutative $U(N)$ gauge theory, but it can also be described by
ordinary gauge theory. Both descriptions
are related by the Seiberg-Witten (SW) map~\cite{SW}, which 
expresses noncommutative gauge fields in terms of fields
with ordinary ``commutative'' gauge transformation properties.

Quantum field theory on non-commutative space-time can be studied also
independently of string theory. 
There are two major approaches. The original one based on actions
that resembles that of Yang-Mills theory with matrix multiplication replaced
by the Moyal-Weyl star product and a more recent one that utilizes the
so-called Seiberg-Witten map to express non-commutative fields in terms
of physical (commutative) fields. Both have their advantages and
limitations. In the original approach unusual non-perturbative effects
like UV/IR mixing~\cite{UVIR} can be studied, but gauge theories are limited to 
the gauge group $U(N)$ in the fundamental representation. There are also
indications of more fundamental problems in the rigorous definition of
the $S$-matrix. The second approach treats non-commutativity 
strictly perturbatively via Seiberg-Witten map expansion in terms of 
$\theta$. A major advantage of the second approach is that models
with any gauge group -- including the one of the standard model -- and 
any particle content can be constructed.
Further problems that are solved in this approach include
the charge quantization problem of NC Abelian gauge theories
and the construction of covariant Yukawa couplings.
The action is manifestly gauge invariant.
It is written in terms of physical fields and their derivatives
and should be understood as an effective model describing non-commutative
effects in particle physics, see \cite{WESS,Zumino,WESS1,Martin} and references
therein.

Experimental signatures of noncommutativity have been discussed from the point of 
view of collider physics~\cite{AY,HPR,HK,MPR} as 
well as low-energy non-accelerator experiments~\cite{MPR,ABDG,CCL}. 
Two widely disparate sets of bounds on $\Lambda_{NC}$
can be found in the literature: bounds of order $10^{11}$ $GeV$~\cite{ABDG} 
or higher~\cite{MPR}, and 
bounds of a few TeV's from colliders~\cite{AY,HPR,HK}.
All these limits rest on one or more of the following assumptions which 
may have to be modified: 
(1) $\theta$ is constant across distances that are very large 
compared with the NC scale; 
(2) unrealistic gauge groups; 
(3) noncommutativity down to low energy scales.
The decay of the $Z$-boson into two photons 
was previously considered in \cite{MPR},
where the authors rely on a noncommutative $U(1)$ model, i.e.,
not yet a bonafide noncommutative model of the electroweak
sector or the standard model.

There are two essential points in which NC gauge  theories differ 
from standard gauge theories. 
The first point is the breakdown of Lorentz invariance with
respect to a fixed non-zero $\theta^{\mu\nu}$ background 
(which obviously fixes preferred directions) and the other
is the appearance of new interactions 
(three-photon coupling, for example) and the modification of standard ones. 
Both properties have a common origin and appear in a number of phenomena.

The action of NC gauge theory
resembles that of ordinary Yang-Mills theory, but with
star products in addition to ordinary matrix multiplication.
The general form of the gauge-invariant action 
for gauge fields is \cite{WESS1}
\begin{equation}
S_{gauge} =-\frac{1}{2}\int d^4x {\bf Tr}\frac{1}{{\bf G}^2}
{\widehat F}_{\mu\nu} \star {\widehat F}^{\mu\nu}.
\end{equation}
Here ${\bf Tr}$ is a trace 
and ${\bf G}$ is an operator that encodes the
coupling constants of the theory. Both will be discussed in detail below.
The NC field strength is 
\begin{equation}
{\widehat F}_{\mu\nu} = \partial_{\mu} {\widehat V}_{\nu} - \partial_{\nu} {\widehat V}_{\mu}
- i[{\widehat V}_{\mu}\stackrel{\star}{,}{\widehat V}_{\nu}]
\end{equation}
and ${\widehat V}_{\mu}$ is the NC analog of the gauge vector potential. 
The Seiberg-Witten maps are used to express the noncommutative fields and 
parameters as functions of 
ordinary fields and parameters and their derivatives. 
This automatically ensures a restriction to the correct degrees of freedom.
For the NC vector potential the SW map yields
\begin{equation}
{\widehat V}_{\xi}=V_{\xi}+\frac{1}{4}{\theta}^{\mu\nu}
\{V_{\nu},(\partial_{\mu}V_{\xi}+F_{\mu\xi})\}+{\cal O}\left(\theta^2 \right),
\end{equation}
where $F_{\mu\nu}\equiv \partial_{\mu}V_{\nu} - \partial_{\nu}V_{\mu} - i[V_{\mu},V_{\nu}]$
is the ordinary field strength and
$V_{\mu}$ is the whole gauge potential for the 
gauge group $G_{SM}\equiv SU(3)_C \times SU(2)_L \times U(1)_Y$
\begin{equation}
V_{\mu}=g'{\cal A}_{\mu}(x)Y + g\sum^3_{a=1}B_{\mu,a}(x)T^a_L + g_s\sum^8_{b=1}G_{\mu,b}(x)T^b_S.
\end{equation}
It is important to realize that the choice of the representation
in the definition of the trace ${\bf Tr}$
has a strong influence on the theory in the noncommutative case.
The reason for this is, that owing to the Seiberg-Witten map, terms
of higher than quadratic order in the Lie algebra generators will 
appear in the trace.
The choice of the trace corresponds to a choice of the representation
of the gauge group. 
The adjoint representation would not lead to 
new triple gauge boson interactions and, in particular, show no
triple-photon vertices~\cite{WESS1,ASCH}.
This, however, would  be an ad hoc choice (unless we are discussing
a GUT scenario.)
Let us emphasize again that the action that we present here
should be understood as an effective theory. 
From this point of view, all representations
of gauge fields that appear in the SM have to be considered in
the definition of the trace. 
Consequently, according to \cite{WESS1}, we choose a trace over all particles 
with different quantum numbers in the model that have
covariant derivatives acting on them.
In the SM, these are, for each generation, five multiplets
of fermions and one Higgs multiplet. 
The operator ${\bf G}$, which determines the coupling constants of the theory,
must commute with all generators 
$(Y,T^a_L,T^b_S)$ of the gauge group,
so that it does not spoil the trace property of ${\bf Tr}$. 
This implies that ${\bf G}$ 
takes on constant values $g_1,\ldots,g_6$
on the six multiplets (Table 1 in Ref.~\cite{WESS1}).
The operator ${\bf G}$ is in general a function of $Y$ 
and the casimirs of $SU(2)$ and $SU(3)$.
However, because of the special assignment of hypercharges in the
SM it is possible to express ${\bf G}$ solely in terms of $Y$.

The action up to linear order in $\theta$ 
allows new triple gauge boson interactions that are forbidden in the SM
and has the following form
\begin{eqnarray}
\lefteqn{S_{gauge}=-\frac{1}{4}\int \hspace{-1mm}d^4x\, f_{\mu \nu} f^{\mu \nu}}
 \label{action2} \\
& &\hspace{-5mm}{}
-\frac{1}{2}\int \hspace{-1mm}d^4x\, {\rm Tr}\left( F_{\mu \nu} F^{\mu \nu}\right)
-\frac{1}{2}\int\hspace{-1mm} d^4x\, {\rm Tr}\left( G_{\mu \nu} G^{\mu \nu}\right)
\nonumber \\
& &\hspace{-5mm}{}
+g_s \,\theta^{\rho\tau}\hspace{-2mm}
\int\hspace{-1mm} d^4x\, {\rm Tr}
\left(\frac{1}{4} G_{\rho \tau} G_{\mu \nu} - G_{\mu \rho} G_{\nu \tau}\right)G^{\mu \nu}\nonumber \\
& &\hspace{-5mm}{}+{g'}^3\kappa_1{\theta^{\rho\tau}}\hspace{-2mm}\int \hspace{-1mm}d^4x\,
\left(\frac{1}{4}f_{\rho\tau}f_{\mu\nu}-f_{\mu\rho}f_{\nu\tau}\right)f^{\mu\nu}
 \nonumber \\
& &\hspace{-5mm}{}+g'g^2\kappa_2 \, \theta^{\rho\tau}\hspace{-2mm}\int
\hspace{-1mm} d^4x \sum_{a=1}^{3}
\left[(\frac{1}{4}f_{\rho\tau}F^a_{\mu\nu}-
f_{\mu\rho}F^a_{\nu\tau})F^{\mu\nu,a}\!+c.p.\right]
 \nonumber \\
& &\hspace{-5mm}{}+g'g^2_s\kappa_3\, \theta^{\rho\tau}\hspace{-2mm}\int
\hspace{-1mm} d^4x \sum_{b=1}^{8}
\left[(\frac{1}{4}f_{\rho\tau}G^b_{\mu\nu}-
f_{\mu\rho}G^b_{\nu\tau})G^{\mu\nu,b}\!+c.p.\right], \nonumber 
\end{eqnarray}
where $c.p.$ means cyclic permutations in $f$.
Here $f_{\mu\nu}$, $F^a_{\mu\nu}$, and $G^b_{\mu\nu}$ are the physical field strengths corresponding 
to the groups $U(1)_Y$, $SU(2)_L$, and $SU(3)_C$, respectively. 
The constants $\kappa_1$, $\kappa_2$, and $\kappa_3$ are parameters of the model.
They are functions of $1/g_i^2,\; (i=1,...6)$ 
and have the following form:
\begin{eqnarray}
\kappa_1 &=& -\frac{1}{g^2_1}-\frac{1}{4g^2_2}+\frac{8}{9g^2_3}-\frac{1}{9g^2_4}+\frac{1}{36g^2_5}
+\frac{1}{4g^2_6},
\nonumber \\
\kappa_2 &=& -\frac{1}{4g^2_2}+\frac{1}{4g^2_5}+\frac{1}{4g^2_6},
\nonumber \\
\kappa_3 &=& +\frac{1}{3g^2_3}-\frac{1}{6g^2_4}+\frac{1}{6g^2_5}.
\end{eqnarray}
In order to match the SM action at zeroth order in $\theta$, three consistency conditions
have been imposed in (\ref{action2}):
\begin{eqnarray}
\frac{1}{{g'}^2} &=& \frac{2}{g^2_1}+\frac{1}{g^2_2}+\frac{8}{3g^2_3}+\frac{2}{3g^2_4}+\frac{1}{3g^2_5}
+\frac{1}{g^2_6},
\nonumber \\
\frac{1}{g^2}&=& \frac{1}{g^2_2}+\frac{3}{g^2_5}+\frac{1}{g^2_6},\nonumber \\
\frac{1}{g_s^2}&=& \frac{1}{g^2_3}+\frac{1}{g^2_4}+\frac{2}{g^2_5}.
\end{eqnarray}
These three conditions together with the requirement that 
$1/g_i^2 > 0$, define a three-dimensional simplex in
the six-dimensional moduli space spanned by $1/g_1^2,...,1/g_6^2$~\footnote{In terms of
the couplings $g_i$ these are complicated equations describing a family of hyper ellipsoids,
however, in terms of $1/g_i^2$ they form a set of linear equations.}.

From the action (\ref{action2}) we extract the 
neutral triple-gauge boson terms which are not present in the SM Lagrangian. 
In terms of physical fields ($A,Z,G$) they are
\begin{eqnarray}
{\cal L}_{\gamma\gamma\gamma}&=&\frac{e}{4} \sin2{\theta_W}\;{\rm K}_{\gamma\gamma\gamma}
{\theta^{\rho\tau}}A^{\mu\nu}\left(A_{\mu\nu}A_{\rho\tau}-4A_{\mu\rho}A_{\nu\tau}\right),\nonumber\\
{\rm K}_{\gamma\gamma\gamma}&=&\frac{1}{2}\; gg'(\kappa_1 + 3 \kappa_2);  \label{L1}\\
& & \nonumber \\
{\cal L}_{Z\gamma\gamma}&=&\frac{e}{4} \sin2{\theta_W}\,{\rm K}_{Z\gamma \gamma}\,
{\theta^{\rho\tau}}
\left[2Z^{\mu\nu}\left(2A_{\mu\rho}A_{\nu\tau}-A_{\mu\nu}A_{\rho\tau}\right)\right.\nonumber\\
& & +\left. 8 Z_{\mu\rho}A^{\mu\nu}A_{\nu\tau} - Z_{\rho\tau}A_{\mu\nu}A^{\mu\nu}\right], \nonumber \\
{\rm K}_{Z\gamma\gamma}&=&\frac{1}{2}\; \left[{g'}^2\kappa_1 + \left({g'}^2-2g^2\right)\kappa_2\right]; \label{L2}\\
& &\nonumber \\
{\cal L}_{ZZ\gamma}&=&{\cal L}_{Z\gamma\gamma}(A\leftrightarrow Z),\nonumber \\
{\rm K}_{ZZ\gamma}&=&\frac{-1}{2gg'}\; \left[{g'}^4\kappa_1 + g^2\left(g^2-2{g'}^2\right)\kappa_2\right]; \label{L3}\\
& &\nonumber \\
{\cal L}_{ZZZ}&=&{\cal L}_{\gamma\gamma\gamma}(A\to Z),\nonumber\\
{\rm K}_{ZZZ}&=&\frac{-1}{2g^2}\; \left[{g'}^4\kappa_1 + 3g^4\kappa_2\right]; \label{L4}\\
& &\nonumber \\
{\cal L}_{Zgg}&=&{\cal L}_{Z\gamma\gamma}(A\to G^b), \nonumber \\
{\rm K}_{Zgg}&=&\frac{g^2_s}{2} \left[1+(\frac{{g'}}{g})^2\right]\kappa_3; \label{L5}\\
& &\nonumber \\
{\cal L}_{\gamma gg}&=&{\cal L}_{Zgg}(Z\rightarrow A), \nonumber \\
{\rm K}_{\gamma gg}&=&\frac{-g^2_s}{2}\;
\left[\frac{g}{g'}+\frac{g'}{g}\right]\kappa_3, \label{L6}
\end{eqnarray} 
where $A_{\mu\nu} \equiv \partial_{\mu}A_{\nu} -
\partial_{\nu}A_{\mu}$, ect.

\begin{figure}
 \resizebox{0.45\textwidth}{!}{%
  \includegraphics{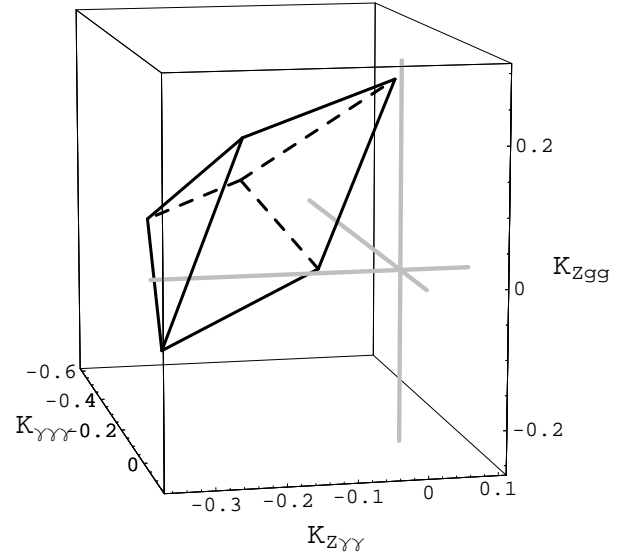}}
 \caption{The three-dimensional simplex that bounds possible values
 for the coupling constants ${\rm K}_{\gamma\gamma\gamma}$, 
 ${\rm K}_{Z\gamma\gamma}$ 
 and ${\rm K}_{Zgg}$ at the  $M_Z$ scale. The vertices of the simplex are:\\
 ($-0.184$,\,$-0.333$,\,$0.054$), ($-0.027$,\,$-0.340$,\,$-0.108$),\\
 ($0.129$,\,$-0.254$,\,$0.217$),
 ($-0.576$,\,$0.010$,\,$-0.108$),\\ ($-0.497$,\,$-0.133$,\,$0.054$), and 
 ($-0.419$,\,$0.095$,\,$0.217$).}
 \label{fig1}
\end{figure}

Fig.(\ref{fig1}) shows the three-dimensional simplex that bounds 
allowed values for the dimensionless coupling constants
${\rm K}_{\gamma\gamma\gamma}$, ${\rm K}_{Z\gamma\gamma}$ 
and ${\rm K}_{Zgg}$. For any choosen point within simplex 
in Fig.(\ref{fig1}) the remaining three coupling constants (\ref{L3},\ref{L4},\ref{L6}), i.e.
${\rm K}_{Z Z \gamma}$, ${\rm K}_{Z Z Z}$ 
and ${\rm K}_{\gamma g g}$ respectively, are uniquely fixed by the NCSM.
This is true for any combination of three coupling constants from equations (\ref{L1}) to (\ref{L6}).

Experimental evidence for noncommutativity coming from the gauge sector, should be searched for 
in processes which involve the  above vertices. 
The simplest and most natural choice are the 
$Z\rightarrow \gamma\gamma, \;gg$ decays, allowed for real (on-shell) particles.
All other simple processes, such as
$\gamma \rightarrow \gamma \gamma, \;gg$, and $Z\rightarrow Z\gamma, \;ZZ$,
are on-shell forbidden by kinematics. 
The $Z\rightarrow \gamma\gamma, \: gg$ decays are strictly
forbidden in the SM by angular momentum conservation and Bose statistics
(Yang Theorem) \cite{Yang,Hagi}, therefore
they both could serve as a clear signal for the existence of 
spacetime noncommutativity~\footnote{$Z$ and $\gamma$ self-couplings
vanish identically in the SM if all particle are on-shell. They can, however, appear if one of 
the photons is considered an off-shell particle in the $s$-channel~\cite{Hagi}.}.

The $Z\rightarrow \gamma\gamma$ process has a tiny SM 
background from the rare $Z\rightarrow \pi^0\gamma,\;\eta\gamma$ decays. 
At high energies, the two photons from the $\pi^0$ or $\eta$ decay
are too close to be separated and they are seen in the electromagnetic calorimeter as 
a single high-energy photon \cite{EXP}. The SM
branching ratios for these rare decays are of order $10^{-11}$
to $10^{-10}$ \cite{ALT}. This is much smaller than the experimental upper bounds
which are of order $10^{-5}$ for the all three branching ratios 
($Z\rightarrow \gamma\gamma,\; \pi^0\gamma,\; \eta\gamma$) \cite{rpp}. 
The experimental upper bound, obtained from the $e^+e^-\rightarrow \gamma\gamma$ annihilation, 
for $\Gamma_{Z \rightarrow \gamma\gamma}$ is $< 1.3\times 10^{-4} GeV$ \cite{rpp}.

The $Z\rightarrow gg$ decay mode should be observed in $Z\rightarrow 2\;{\rm jets}$ processes.
However, it could be smothered by the strong 
$Z\rightarrow q{\bar q}$ background, i.e. by hadronization, which also
contains NC contributions. Since
the hadronic width of the $Z$ is in good agreement with the QCD corrected SM, 
the $Z\rightarrow gg$ can at most be a few percent.
Taking into account the discrepancy between the experimentally 
observed hadronic width for the $Z$-boson 
and the theoretical estimate based on the radiatively corrected SM, 
we estimate the upper bound for any new hadronic
mode, like $\Gamma_{Z \rightarrow gg}$, to be $\sim 10^{-3}\; GeV$ \cite{rpp}.

We now derive the partial widths for the $Z(p) \rightarrow \gamma (k)\,\gamma (k')$ decay.
Care has to be taken when one tries to compute matrix elements in NCGFT. In our model, the 
\emph{in} and \emph{out} states can be taken to be ordinary \emph{commutative} particles.
Quantization is straightforward to the order in $\theta$ that we have considered;
Feynman rules can be obtained either via the Hamiltonian formulation or directly from the
Lagrangian; a rather convenient property of the action, relevant to computations, is
its symmetry under ordinary gauge transformations in addition to noncommutative ones.
From the Lagrangian ${\cal L}_{Z\gamma \gamma}$, it is easy to write 
the gauge-invariant amplitude ${\cal M}_{Z\rightarrow \gamma\gamma}$ in momentum space.
Since we are dealing with a SM forbidden process, 
this is essentially done using distorted wave Born approximation.
It gives: 
\begin{eqnarray}
\sum_\mathrm{spins}\,|{\cal M}_{Z\rightarrow \gamma \gamma}|^2 
= -{\theta}^2 + \frac{8}{M^2_Z}(p{\theta}^2 p)
- \frac{16}{M^4_Z}(k{\theta}k')^2 \, .
\label{eqn0}
\end{eqnarray}
From above equation and in the $Z$-boson rest frame, the partial width
of the $Z \rightarrow \gamma\gamma$ decay is
\begin{equation}
\Gamma_{Z\rightarrow \gamma\gamma} 
= \frac{\alpha}{12} M^5_Z \sin^2 2\theta_W {\rm K}^2_{Z\gamma \gamma} 
\left[\frac{7}{3}({\vec {\theta}}_T)^2+({\vec {\theta}}_S)^2\right],
\label{eqn1}
\end{equation}
where ${\vec {\theta}}_T=\{{\theta^{01}},{\theta^{02}},{\theta^{03}}\}$ 
and ${\vec {\theta}}_S=\{{\theta^{23}},{\theta^{13}},{\theta^{12}}\}$, are
responsible for time-space and space-space noncommutativity, respectively. 
This result differs essentially from that given in \cite{MPR}
where the $\Gamma_{Z\rightarrow \gamma\gamma}$
partial width depends only on time-space noncommutativity.

For the $Z$-boson at rest and polarized in the 
direction of the $3$-axis, we find that the \emph{polarized} partial width is
\begin{eqnarray}
& &\Gamma_{Z^3 \rightarrow \gamma\gamma }\;=\;
\frac{\alpha}{4} \;M^5_Z \;\sin^2 2\theta_W\;{\rm K}^2_{Z\gamma \gamma} 
\nonumber \\
& &\times \left[\frac{2}{5}
\left(({\theta}^{01})^2+({\theta}^{02})^2\right)
+\frac{23}{15}({\theta}^{03})^2+({\theta}^{12})^2\right]. \label{eqn2}
\end{eqnarray}
In the absence of time-space noncommutativity 
a sophisticated, sensibly arranged  polarization
experiment could in principal determine the vector of ${\vec {\theta}}_S$. 
A NC structure of spacetime may depend on the matter that is present. 
In our case it is conceivable that the direction of ${\vec {\theta}}_{T,S}$
may be influenced by the polarization of the $Z$ particle.
In this case, our result for the \emph{polarized} partial width is particularly relevant.

Due to the same Lorentz structure of
the Lagrangians ${\cal L}_{Z\gamma\gamma}$ and ${\cal L}_{Zgg}$ we find
\begin{eqnarray}
\frac{\Gamma_{Z\rightarrow gg}}{\Gamma_{Z\rightarrow \gamma\gamma}}\;=\;
\frac{\Gamma_{Z^3\rightarrow gg}}{\Gamma_{Z^3\rightarrow \gamma\gamma}}\;=\;
8\frac{{\rm K}^2_{Zgg}}{{\rm K}^2_{Z\gamma \gamma}}. \label{eqn3}
\end{eqnarray}
The factor of eight in the above ratios is due to color.

In order to estimate the NC parameter from upper bounds 
$\Gamma^{exp}_{Z \rightarrow \gamma\gamma} < 1.3 \times 10^{-4} GeV$ and
$\Gamma^{exp}_{Z \rightarrow gg} < 1 \times 10^{-3}\; GeV$ \cite{rpp}
it is necessary to determine the range of couplings ${\rm K}_{Z\gamma\gamma}$ and ${\rm K}_{Zgg}$.
\begin{figure}
 \resizebox{0.45\textwidth}{!}{%
  \includegraphics{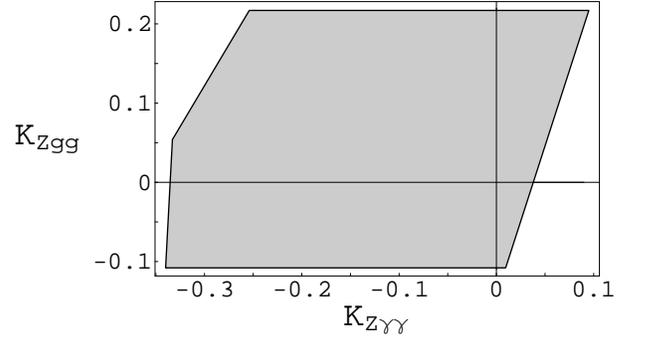}}
 \caption{The allowed region for ${\rm K}_{Z\gamma\gamma}$ and ${\rm K}_{Zgg}$ 
at the $M_Z$ scale, projected from the simplex given in Fig~1. 
Note, that ${\rm K}_{\gamma\gamma\gamma}$          is nonzero at the point where both
${\rm K}_{Z\gamma\gamma}$ and ${\rm K}_{Zgg}$  vanish.
The vertices of the polygon are:
$(-0.254,\, 0.217)$, 
$(-0.333,\, 0.054)$, 
$(-0.340,\, -0.108)$, 
$(0.010,\, -0.108)$ and 
$(0.095, \,0.217)$.}
 \label{fig2a}
\end{figure}
The allowed region for coupling constants ${\rm K}_{Z\gamma\gamma}$ and ${\rm K}_{Zgg}$ 
is given in Fig.(\ref{fig2a}).
Since ${\rm K}_{Z\gamma\gamma}$ and ${\rm K}_{Zgg}$ could be zero
simultaneously it is not possible to extract an upper bound on $\theta$ 
only from the above experimental upper bounds alone.

We  need to consider an extra interaction from the NCSM gauge sector, namly the
 triple photon vertex, to estimate $\theta$.
\begin{figure}
 \resizebox{0.45\textwidth}{!}{%
  \includegraphics{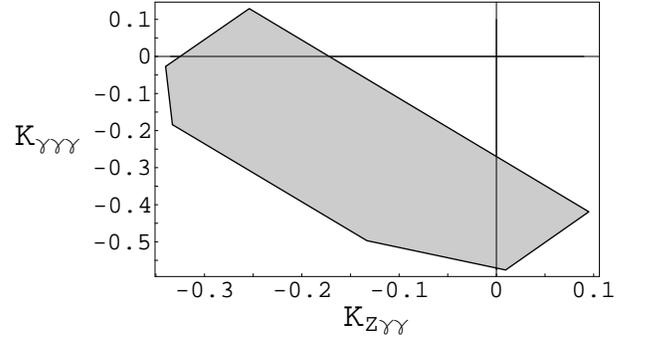}}
 \caption{The allowed region for ${\rm K}_{Z\gamma\gamma}$ 
 and ${\rm K}_{\gamma\gamma\gamma}$  at the $M_Z$ scale, projected from
 the simplex given in Fig 1. The vertices of the polygon are:
 $(-0.333,\, -0.184)$, $(-0.340,\, -0.027)$, $(-0.254,\, 0.129)$, $(0.095,\, -0.419)$, 
 $(0.0095, \,-0.576)$, and $(-0.133,\, -0.497)$.}
 \label{fig2b}
\end{figure}
The important point is that the triplet of
coupling constants 
${\rm K}_{\gamma\gamma\gamma}$, ${\rm K}_{Z\gamma\gamma}$ and ${\rm K}_{Zgg}$,
as well as the pair of couplings ${\rm K}_{\gamma\gamma\gamma}$ and ${\rm K}_{Z\gamma\gamma}$
{\it can never vanish simultaneously} due to the constraint set by the
value of the SM coupling constants at the weak interaction scale. This can be seen
from the simplex in Fig.(\ref{fig1}). 
In conclusion, it is possible to estimate $\theta$ 
from the NCSM gauge sector through a combination of various types of
processes containing the $\gamma\gamma\gamma$ and $Z\gamma\gamma$ vertices.
These are processes of the type $2 \rightarrow 2$, such as 
$e^+e^-\rightarrow \gamma\gamma$, 
$e\gamma \rightarrow e\gamma$, and $\gamma\gamma \rightarrow e^+e^-$ 
in leading order.
The analysis has to be carried out in the same way as in Ref.\cite{HPR}.
Theoreticaly consistent modifications of relevent vertices are, however, necessary. 
\begin{figure}
 \resizebox{0.45\textwidth}{!}{%
  \includegraphics{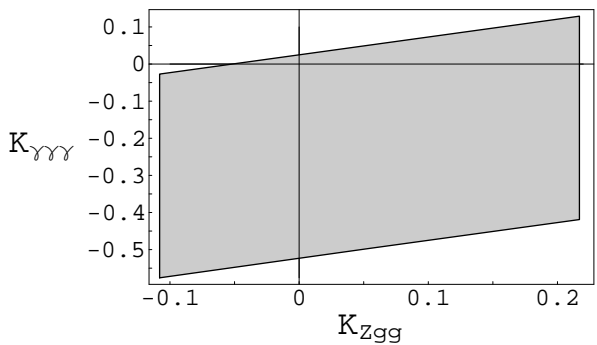}}
 \caption{The allowed region for ${\rm K}_{\gamma\gamma\gamma}$
 and ${\rm K}_{Zgg}$ at the $M_Z$ scale, projected from the simplex
 given in Fig~1. Note, that ${\rm K}_{Z\gamma\gamma}$ is non-zero at
the point where both ${\rm K}_{\gamma\gamma\gamma}$
 and ${\rm K}_{Zgg}$ vanish. 
The vertices of the polygon are: $(-0.108,\, -0.576)$, $(-0.108,\, -0.027)$, 
 $(0.217,\, 0.129)$, $(0.217,\, -0.419)$, and $(0.054,\, -0.497)$.}
 \label{fig2c}
\end{figure}
Finally, we present the allowed region for pair of couplings ${\rm K}_{\gamma\gamma\gamma}$ and
${\rm K}_{Zgg}$ in Fig.(\ref{fig2c}). 
Note, that
Figs.(\ref{fig2a}) to (\ref{fig2c}) represent projections of pairs of coupling constants
from the three dimensional simplex spanned by the constants 
${\rm K}_{\gamma\gamma\gamma}$, ${\rm K}_{Z\gamma\gamma}$ and ${\rm K}_{Zgg}$.

The structure of our main results (\ref{eqn0}) to (\ref{eqn3}) remains the same
for  $SU(5)$ and $SU(3)_C \times SU(3)_L \times SU(3)_R$ GUT's that
embed the NCSM that is based on the SW map \cite{Desh,ASCH}; only the coupling
constants change. Note, in the particular case of $SO(10)$ GUT
there is no triple gauge boson coupling \cite{ASCH}.

In this article we have propose two SM strictly forbidden decay modes, namely,
$Z \rightarrow \gamma\gamma, gg$, as a possible signature of the NCSM. 
An experimental discovery of $Z \rightarrow \gamma\gamma, gg$
decays would certainly indicate a violation of the presently accepted SM and definitive apperance
of new physics. 

In conclusion, the gauge sector of the nonminimal NCSM is an excellent place to discover
spacetime noncommutativity experimentally~\cite{LEPEWWG,L3,DELPHI,OPAL,muller}, but not the best place to find bounds that exclude it.
We hope that the importance of a possible
discovery of noncommutativity of spacetime will convince
experimentalists to look for SM forbidden decays in the gauge sector.
A good reason for this is that the sensitivity to 
the noncommutative parameter $\theta^{\mu\nu}$ could be in a range of the next
generation of linear colliders with a c.m.e. around a few TeV's. 
\\
\begin{acknowledgments}
We would like to thank P.\ Aschieri, B.\ Jur\v co and H.~\v Stefan\v ci\' c for helpful discussions.
One of us (NGD) 
would like to thank the University of Hawaii Theory Group for hospitality.
This work was supported by the Ministry of Science and 
Technology of the Republic of Croatia under Contract No. 0098002, and by the 
US Department of Energy, Grant No.\ 
DE-FG06-85ER 40224.
\end{acknowledgments}


\end{document}